\documentclass[
    ,final            
  ]
  {aipproc}

\layoutstyle{6x9}

\begin{document}

\newcommand{\gsim}{\mbox{${~\raise.25em\hbox{$>$}\kern-.70em
\lower.25em\hbox{$\sim$}~}$}}
\newcommand{\lsim}{\mbox{${~\raise.25em\hbox{$<$}\kern-.70em
\lower.25em\hbox{$\sim$}~}$}}

\title{Topics in Leptogenesis\footnote{Based on work done in collaboration
    with G. Engelhard, Y. Grossman, Y. Nir, J. Racker and E. Roulet.  } }


\author{Enrico Nardi}{
  address={INFN, Laboratori Nazionali di Frascati, C.P. 13,
      100044 Frascati, Italy\\ and \\
    Instituto de F\'{i}sica, Universidad de Antioquia,
    A.A.1226, Medell\'{i}n, Colombia}
}

\begin{abstract}
  Baryogenesis via leptogenesis provides an appealing mechanism to explain the
  observed baryon asymmetry of the Universe.  Recent refinements in the
  understanding of the dynamics of leptogenesis include detailed studies of
  the effects of lepton flavors and of the role possibly played by the lepton
  asymmetries generated in the decays of the heavier singlet neutrinos
  $N_{2,3}$. In this talk I present a short  review of these recent
  developments in the theory of leptogenesis.
\end{abstract}

\maketitle


\section{Introduction}
The possibility that the Cosmic Baryon Asymmetry (BAU) could originate from a
lepton number asymmetry generated in the $CP$ violating decays of the heavy
seesaw Majorana neutrinos was put forth twenty years ago by Fukugita and
Yanagida~\cite{Fukugita:1986hr}.  Their proposal came shortly after Kuzmin,
Rubakov and Shaposhnikov pointed out that above the electroweak phase
transition $B+L$ is violated by fast electroweak anomalous
interactions~\cite{Kuzmin:1985mm}.  This implies that any lepton asymmetry
would be in part converted into a baryon asymmetry. However, the discovery
that at $T\gsim 100\,$GeV electroweak interactions do not conserve baryon
number, also suggested the exciting possibility that baryogenesis could be a
purely standard model phenomenon, and opened the way to electroweak
baryogenesis~\cite{Farrar:1993sp}.  Indeed, in the early 90's electroweak
baryogenesis attracted more interest than leptogenesis. Still, a few
remarkable papers appeared that put the first basis for {\it quantitative}
studies of leptogenesis.  Here I will just mention two important
contributions: the 1992 Luty's paper~\cite{Luty:1992un} in which the rates for
several processes relevant for the Boltzmann equations for leptogenesis were
first presented, and the 1996 paper of Covi, Roulet and
Vissani~\cite{Covi:1996wh} that computed the correct expression for the $CP$
asymmetry in the decays of the lightest Majorana neutrino.

Around year 2000 a flourishing of detailed studies of leptogenesis begins,
with a corresponding burst in the number of papers dealing with this subject
(see fig.~1). This raise of interest in leptogenesis can be traced back to two
main reasons: firstly, by this time it became clear that within the standard
model, electroweak baryogenesis fails to reproduce the correct BAU by many
orders of magnitude, and that even in supersymmetric models this scenario
quite likely is not viable.  Secondly, the experimental confirmation (from
oscillation experiments) that neutrinos have nonvanishing masses strengthened
the case for the seesaw mechanism, that in turn implies the existence, at some
large energy scale, of lepton number violating ($\not \!  L$)
interactions.

\begin{figure}  
\includegraphics[height=.32\textheight]{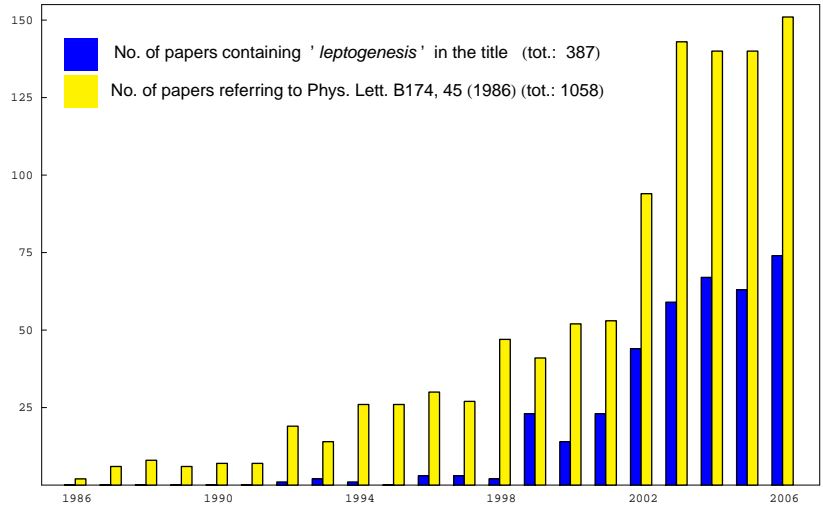}
  \caption{Barchart showing   
    the number of publications in each year containing a reference to the
    Fukugita and Yanagida paper \cite{Fukugita:1986hr} (yellow) and
    containing the word 'leptogenesis' in the title (blue).  }
\label{fig1}
\end{figure}

The number of important papers and the list of people that contributed to the
development of leptogenesis studies and to understand the various implications
for the low energy neutrino parameters is too large to be recalled here. Let
me just mention the remarkable paper of Giudice {\it et
  al.}~\cite{Giudice:2003jh} that appeared in 2003: in this paper a whole set
of thermal corrections for the relevant leptogenesis processes were carefully
computed, a couple of mistakes common to previous studies were pointed out and
corrected, and a detailed numerical analysis was presented both for the SM and
the MSSM cases.  Eventually, it was claimed that the residual numerical
uncertainties would probably not exceed the 10\%-20\%
level~\cite{Giudice:2003jh}.  A couple of years later, Nir, Roulet, Racker and
myself~\cite{Nardi:2005hs} carried out a detailed study of additional effects
that were not accounted for in the analysis of ref.~\cite{Giudice:2003jh}.
This included electroweak and QCD sphaleron effects, the effects of the
asymmetry in the Higgs number density, as well as the constraints on the
particles asymmetry-densities implied by the spectator reactions that are in
thermal equilibrium in the different temperature ranges relevant for
leptogenesis~\cite{Nardi:2005hs}.  Indeed, we found that the largest of theses
new effects would barely reach the level of a few tens of percent.

However, two important ingredients that had been overlooked in practically all
previous leptogenesis studies, had still to be accounted for.  These were the
role of the light lepton flavors, and the role of the heavier seesaw Majorana
neutrinos.  One remarkable exception was the 1999 paper of Barbieri {\it et
  al.}~\cite{Barbieri:1999ma} that, besides addressing as the main topic the
issue of flavor effects in leptogenesis, pointed out that the lepton number
asymmetries generated in the decays of the heavier seesaw neutrinos should
also be taken into account in computing the BAU.\footnote{Lepton flavor
  effects were also considered by Endoh, Morozumi and Xiong in their 2003
  paper~\cite{Endoh:2003mz}, in the context of the minimal seesaw model with
  just two right handed neutrinos.}  However, these important results did not
have much impact on subsequent analyses. The reason might be that these were
thought to be just order one effects on the final value of the BAU, with no
other major consequences for leptogenesis. As I will discuss in the following,
the size of the effects 
could easily reach the level of one order of magnitude, and, most importantly,
they can spoil the leptogenesis constraints on the neutrino low energy
parameters, and in particular the limit on the absolute scale of neutrino
masses~\cite{buch02-03}. This is important, since it was thought that this
limit was a firm prediction of leptogenesis with hierarchical seesaw
neutrinos, and that the discovery of a neutrino mass $m_\nu\gsim 0.2\,$eV
would have strongly disfavored leptogenesis, or hinted to different scenarios
(as e.g. resonant leptogenesis~\cite{pila0405}).

\section{The standard scenario}

Let us start by writing the first few terms of the leptogenesis Lagrangian:
\begin{equation}
\label{lagrangian1}
{\cal L} =\frac{1}{2}\left[\bar N_1 (i\! \not\! \partial) N_1 - 
M_1 N_1 N_1\right] -(\lambda_{1}\,
\bar N_1\,  \ell_{1}{H} +{\rm h.c.}).
\end{equation}
Here $N_1$ is the lightest right-handed Majorana neutrino with mass
$M_1$, $H$ is the Higgs field, and $\ell_1$ is the lepton doublet to
which $N_1$ couples, that when expressed on a complete  orthogonal basis 
$\{\ell_i \}$  reads 
\begin{equation}
|\ell_1\rangle  = (\lambda  \lambda^\dagger )_{11}^{-1/2} 
\sum_i \lambda_{1i}\,|\ell_i\rangle.
\end{equation}
In practice it is always convenient to use the basis that diagonalizes the
charged lepton Yukawa couplings (the flavor basis) that also has well defined
$CP$ conjugation properties $CP(\{\ell_i \})=\{\bar \ell_i \}$ with
$i=e,\,\mu,\,\tau $.  Note that in the first and third term in
(\ref{lagrangian1}) a lepton number can be assigned to $N_1$, that is however
violated by two units by the mass term.  Then eq.~(\ref{lagrangian1}) implies
processes that violate $L$ like inverse-decays followed by $N_1$ decays
$\ell_1 \leftrightarrow N_1 \leftrightarrow\bar \ell_1$ or of-shell $\ell_1
H\leftrightarrow \bar\ell_1 \bar H$ scatterings.  At large temperatures $T\gg
M_1$, $N_1$ decays and inverse decays are blocked because of thermal
effects\footnote{$CP$ violating Higgs decays can kinematically occur when
  $T\gsim 7 M_1$~\cite{Giudice:2003jh} however, because of a large time
  dilation factor, only a small fraction of the Higgses can actually decay
  before thermal phase space closes.}  and off-shell $\not\!\!  L$ processes
are suppressed with respect to $L$-conserving ones as $M_1^2/T^2$.  At $T\ll
M_1$ decays and inverse decays are Boltzmann suppressed, while $ \ell_1
\leftrightarrow \bar \ell_1$ scatterings are suppressed as $T^2/M_1^2$.
Therefore the temperature range in which $\not\!\! L$ processes can be
important for leptogenesis is around $T\sim M_1$. The possibility of
generating an asymmetry between the number of leptons $n_{\ell_1}$ and
antileptons $n_{\bar \ell_1}$ is due to a non-vanishing $CP$ asymmetry in
$N_1$ decays:
\begin{equation}
\epsilon_1\equiv \frac{
\Gamma(N_1\to \ell_1 H)-
\Gamma(N_1\to \bar \ell_1 \bar H)}
{\Gamma(N_1\to \ell_1 H)+
\Gamma(N_1\to \bar \ell_1 \bar H)} \neq 0.
\end{equation}
In order that a macroscopic $L$ asymmetry can build up, the condition
that $\not\!\! L$ reactions are (at least slightly) out of equilibrium at
$T\sim M_1$ must also be satisfied. This condition can be expressed in terms of two
dimensionful parameters, defined in terms of the Higgs vev $v\equiv \langle
H\rangle$ and of the Plank mass $M_P$ as:
\begin{equation}
\label{mstar}
\tilde m_1 =\frac{(\lambda\lambda^\dagger)_{11}\,v^2 }{M_1}, \qquad\qquad \quad  m_* \approx
10^3\,\frac{v^2}{M_P}\approx 10^{-3}\,{\rm eV}.
\end{equation}  
The first parameter ($\tilde m_1$) is related to the rates of $N_1$ processes
(like decays and inverse decays) while the second one ($m_*$) is related to
the expansion rate of the Universe at $T\sim M_1$.  When $\tilde m_1< m_*$
$\not\!\! L$ processes are slower than the expansion and leptogenesis can
occur.  As $\tilde m_1$ increases to values larger than $m_*$, $\not\!\! L$
reactions approach thermal equilibrium thus rendering leptogenesis inefficient
because of the back-reactions that tend to erase any macroscopic asymmetry.
However, even for $\tilde m_1$ as large as $\sim 100\,m_*$ a lepton asymmetry
sufficient to explain the BAU can be generated.  It is customary to refer to
the condition $\tilde m_1> m_*$ as to the {\it strong washout regime} since
washout reactions are rather fast.  This regime is considered more likely than
the {\it weak washout regime} $\tilde m_1< m_*$ in view of the experimental
values of the light neutrino mass-squared differences (that are both $>
m_*^2$) together with the theoretical lower bound $\tilde m_1 \geq m_{\nu_1}$,
where $m_{\nu_1}$ is the mass of the lightest neutrino.  The strong washout
regime is also theoretically more appealing since the final value of the
lepton asymmetry is independent of the particular value of the  $N_1$ initial
abundance and also of a possible asymmetry $Y_{\ell_1}\! =(n_{\ell_1}-n_{\bar
  \ell_1})/s\neq 0$ ($s$ is the entropy density) preexisting the $N_1$ decay
era.  This last fact has been often used to argue that for $\tilde m_1> m_*$
only the dynamics of the lightest Majorana neutrino $N_1$ is important, since
asymmetries generated in the decays of the heavier $N_{2,3}$  would
be efficiently erased by the strong $N_1$-related washouts.  As we will see
below, the effects of $N_1$ interactions on the $Y_{\ell_{2,3}}$ asymmetries
are subtle, and the previous argument is incorrect.
The result of numerical integration of the Boltzmann equations for
 $Y_{\ell_1}$ can be    conveniently expressed in terms of an
efficiency factor $\eta_1$, that  ranges between 0 and 1:
\begin{equation}
Y_{\ell_1} = 3.9\times 10^{-3}\  \eta_1 \epsilon_1, \qquad\qquad \qquad \eta_1 \approx \frac{m_*}{\tilde m_1}.
\end{equation}
The second relation gives a rough approximation for $\eta_1$ in the strong
washout regime that will become useful in analyzing the impact of flavor
effects.  The possibility of deriving an upper limit for the the light
neutrino masses~\cite{buch02-03} follows from the existence of a theoretical
bound on the maximum value of the $CP$ asymmetry $\epsilon_1$ (that holds when
$ N_{1,2,3}$ are sufficiently hierarchical) and relates $M_1$, $m_{\nu_3}$ and
the washout parameter $\tilde m_1$:
\begin{equation}
\label{limit}
  | \epsilon_1| \leq    \left[\frac{3}{16\pi}\frac{M_1}{v^2} (m_{\nu_3}-m_{\nu_1})\right]
  \,\sqrt{1-\frac{m_{\nu_1}^2}{\tilde m_1^2}}. 
\end{equation}
The term in square brackets is the so called Davidson-Ibarra limit \cite{da02}
while the square root is a correction that was first given in \cite{hamby03}.
When $m_{\nu_3}\gsim 0.1\,$eV, the light neutrinos are quasi-degenerate and
$m_{\nu_3}-m_{\nu_1}\sim \Delta m^2_{atm}/2 m_{\nu_3} \to 0$ so that, to keep
$\epsilon_1$ finite, $M_1$ is pushed to large values $\gsim 10^{13}\,$GeV.
Since at the same time $\tilde m_1$ must remain larger than $m_{\nu_{1,3}} $
the washout effects also increase, until the surviving lepton-asymmetry is
too small to explain the BAU.\footnote{$\Delta L=2$ washout processes, that
  depend on a different parameter than $\tilde m_1$, and that can become
  important when $M_1$ is large, also play a role in establishing the limit.}
The  interesting  limit  $m_{\nu_3} \lsim 0.15\,$eV results.
 
\section{Lepton flavor effects}

In the Lagrangian (\ref{lagrangian1}) the terms involving the charged
lepton Yukawa couplings have not been included.  Since all these
couplings are rather small, if leptogenesis occurs at temperatures $T
\gsim 10^{12}\,$GeV, when the Universe is still very young, not many of
the related (slow) processes could have occurred during its short
lifetime, and leptogenesis has essentially no knowledge of lepton
flavors. At $T\lsim 10^{12}\,$GeV the reactions mediated by the tau
Yukawa coupling  $h_\tau$ become important, and at $ T\lsim 10^{9}\,$GeV
also $h_\mu$-reactions have to be accounted for. 
Including the  Yukawa terms for the leptons  yields the  Lagrangian:
\begin{equation}
\label{lagrangian2}
{\cal L} =\frac{1}{2}\left[\bar N_1 (i\! \not\! \partial) N_1 - 
M_1 N_1 N_1\right] -(\lambda_{1i}\,
\bar N_1\,  \ell_{i}{H} +h_i\bar e_i\ell_i H^\dagger +  {\rm h.c.}),
\end{equation}
where, since we are using the flavor basis, the matrix $h$ of the Yukawa
couplings is diagonal.  The flavor content of the (anti)lepton doublets
$\ell_1$ ($\bar \ell'_1$) to which $N_1$ decays is now important, since these
states do not remain coherent, but are effectively resolved into their flavor
components by the fast Yukawa interactions
$h_i$~\cite{Barbieri:1999ma,Abada:2006fw,Nardi:2006fx}. Note that in general,
due to $CP$ violating loop effects, $CP(\bar \ell'_1)\neq \ell_1$, that is the
antileptons produced in $N_1$ decays are not the $CP$ conjugated of the
leptons, implying that the flavor projections $K_i\equiv |\langle\ell_i
|\ell_1 \rangle |^2$ and $\bar K_i\equiv |\langle\bar\ell_i |\bar\ell_1'
\rangle |^2$ differ: $\Delta K_i = K_i-\bar K_i \neq 0$.  The fact that
$\ell_1$ and $\bar \ell'_1$ can differ in their flavor content implies that
even when $N_1$ decays with equal branchings into leptons and antileptons
(yielding $\epsilon_1=0$) the $CP$ asymmetries for the decays into single
flavors can still be non-vanishing. The flavor $CP$ asymmetries 
are defined as~\cite{Nardi:2006fx}:
\begin{equation}
\epsilon_1^i = \frac{
\Gamma(N_1\to \ell_i H)-
\Gamma(N_1\to \bar \ell_i \bar H)}
{\Gamma_{N_1}}=K_i\epsilon_1 + {\Delta K_i}/{2}.
\end{equation}
The factor $\Delta K_i$ in the second equality accounts for the flavor
mismatch between leptons and antileptons, while the factor
$K_i$ in front of $\epsilon_1$ accounts for the reduction in the strength of
the $N_1$-$\ell_i$ coupling with respect to $N_1$-$\ell_1$. The same factor
also reduces the strength of the washouts for the $i$-flavor, yielding an
efficiency factor $\eta_{1}^i=\min (\eta_1/K_i,1)$.
Assuming for illustration  $\eta_1/K_i < 1$ the resulting 
asymmetry is
\begin{equation}
Y_L \approx \sum_i \epsilon^i_1\, \eta_{1}^i \approx n_f 
Y_{\ell_1} + \sum_i\frac{\Delta K_i}{2K_i}\>\frac{m_*}{\tilde m_1}.
\end{equation}
In the first term on the r.h.s. $n_f$ represents the number of flavors
effectively resolved by the charged lepton Yukawa interactions ($n_f=2$ or 3),
while $Y_{\ell_1}$ is the asymmetry that would have been obtained by
neglecting the decoherence of $\ell_1$. The second term, that is controlled by
the 'flavor mismatch' factor $\Delta K_i$, can become particularly large in
the cases when the flavor $i$ is almost decoupled from $N_1$ ($K_i \ll 1$).
This situation is depicted in fig.~2 for the two-flavor case and for two
different temperature regimes.  The two flat curves give $|Y_{B-L}|$ as a
function of the flavor projector $K_\tau$ assuming $\Delta K_\tau=0$, and show
rather clearly the enhancement of a factor $\approx 2$ with respect to the
aligned cases (the points at $K_\tau=0,\,1$) for which flavor effects are
irrelevant. The other two curves, that peak at values close to the boundaries,
when $\ell_\tau$ or a combination orthogonal to $\ell_\tau$ are almost
decoupled from $N_1$, show that $\ell_1$-$\ell'_1$ flavor mismatch can produce
much larger enhancements.  In conclusion, the relevance of flavor effects is
at least twofold:
 
\begin{figure} 
\hspace{-1.9cm}
\includegraphics[width=.35\textheight,height=13truecm,angle=270]{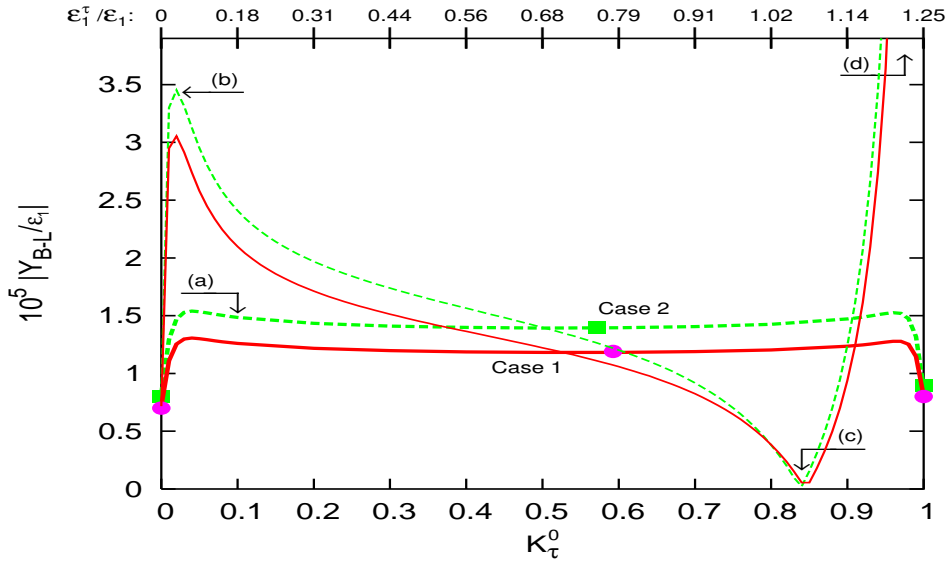}
  \caption{
    $|Y_{B-L}|$ (in units of $10^{-5}|\epsilon_1|$) as a function of $K_\tau $
    in two two-flavor regimes.  The thick lines correspond to the special case
    when $K_\tau =\bar K_\tau$. The thin lines give an example of the
    results for $K_\tau \neq \bar K_\tau$. The filled circles and squares at
    $K_\tau = 0\,,1$ correspond to the aligned cases without flavor effects.}
\end{figure}

\begin{itemize}
\item[1.] 
The  BAU resulting form leptogenesis can be several times larger 
than what would be obtained neglecting flavor effects. 
\item[2.] 
If leptogenesis occurs at temperatures when flavor
effects are important, the limit on the light neutrino 
masses does not hold \cite{Abada:2006fw,DeSimone:2006dd}.   
This is because there is no analogous of the Davidson-Ibarra bound  
eq.~(\ref{limit}) for the flavor asymmetries $\epsilon_1^i$. 

\end{itemize}

\section{The effects of the  heavier Majorana neutrinos}

What about the possible effects of the heavier Majorana neutrinos $N_{2,3}$
that we have so far neglected ?  Some recent studies analyzed the so called
``$N_1$-decoupling'' scenario, in which the Yukawa couplings of $N_1$ are
simply too weak to washout an asymmetry generated in $N_2$ decays (and
$\epsilon_1$ is too small to explain the BAU)~\cite{decoupling}.
This is a consistent scenario in which $N_2$ leptogenesis could successfully
explain the BAU.  However, in the opposite situation when the Yukawa couplings
of $N_1$ are very large, it was generally assumed that the asymmetries related
to $N_{2,3}$ are irrelevant for the computation of the BAU, since they would
be washed out during $N_1$ leptogenesis.  In contrast to this, in
ref.~\cite{Barbieri:1999ma} (and more recently also in
ref.~\cite{Strumia:2006qk}) it was stated that part of the asymmetry from
$N_{2,3}$ decays does in general survive, and must be taken into account when
computing the BAU.  In ref.~\cite{Engelhard:2006yg} Engelhard, Grossman, Nir
and myself carried out a detailed study of the fate of a lepton asymmetry
preexisting $N_1$ leptogenesis, and we reached conclusions that agree with
these statements. I will briefly describe the reasons for this and the
importance of the results.  Including also $N_{2,3}$ the leptogenesis
Lagrangian reads:
\begin{equation}
\label{lagrangian3}
{\cal L} =\frac{1}{2}\left[\bar N_\alpha (i\! \not\! \partial) N_\alpha - 
{N_\alpha} M_\alpha N_\alpha\right] -(\lambda_{\alpha i}\,
\bar N_\alpha\,  \ell_{i}{H} +h_i\bar e_i\ell_i H^\dagger +  {\rm h.c.}),
\end{equation}
where $\alpha=1,2,3$ is a heavy neutrinos index, and the heavy neutrinos are
written in the mass basis.  It is convenient to define the three (in general
non-orthogonal) combinations of lepton doublets $\ell_\alpha$ to which the
corresponding $N_\alpha$ decay:
\begin{equation}
|{\ell_\alpha}\rangle=(\lambda\lambda^\dagger)_{\alpha\alpha}^{-1/2}
\sum_i\lambda_{\alpha i}\,|{\ell_i}\rangle.
\end{equation}
Let us discuss for definiteness the case where a sizeable asymmetry originates
from $N_2$ decays, while $N_1$ leptogenesis is instead rather inefficient.
That is, let us assume that $N_2$-related washouts are not too strong, while
$N_1$-related washouts are so strong that by itself $N_1$ leptogensis would
not be successful:
\begin{equation}\label{conditions} 
\tilde m_2\not\gg m_*,  \qquad\qquad\qquad \tilde m_1\gg m_*.
\end{equation}
To simplify the arguments, let us also impose two additional conditions:
thermal leptogenesis, that is a vanishing initial $N_1$ abundance
$n_{N_1}(T\gg M_1)\approx0$, and a strong hierarchy $M_2/M_1\gg1$. From this
it follows that there are no $N_1$ related washout effects during $N_2$
leptogenesis and, because $n_{N_2}(T\approx M_1)$ is Boltzmann suppressed,
there are no $N_2$ related washouts during $N_1$ leptogenesis. Thus $N_2$ and
$N_1$ dynamics are decoupled. Now, the second condition in (\ref{conditions})
implies that already at $T\gsim M_1$ the interactions mediated by the $N_1$
Yukawa couplings are sufficiently fast to quickly destroy the coherence of the
state $\ell_2$ produced in $N_2$ decays.  Then a statistical mixture of
$\ell_1$ and of the states orthogonal to $\ell_1$ builds up, and it can be
described by a suitable {\it diagonal} density matrix.  On general grounds one
expects that decoherence effects proceed faster than washout. In the relevant
range, $T\gsim M_1$, this is also ensured by the fact that because of thermal
effects the dominant ${\cal O}(\lambda^2)$ washout process (the inverse decay
$\ell H \to N_1$) is blocked \cite{Giudice:2003jh}, and only scatterings with
top-quarks and gauge bosons, that have additional suppression factors of
$h_t^2$ and $g^2$, contribute to the washout.

Let us consider the case where both $N_{2}$ and $N_1$ decay at
$T\gsim10^{12}\,$GeV and flavor effects are irrelevant.  In this regime a
convenient choice for the orthogonal lepton basis is
$(\ell_1,\,\ell_0,\,\ell'_0)$ where, without loss of generality, $\ell'_0$ can
be chosen to satisfy $\langle\ell'_0|\ell_2\rangle=0$.  Then the asymmetry
$Y_{\ell_2}$ produced in $N_2$ decays decomposes in the two components:
\begin{equation}
\label{eq:c2}
Y_{\ell_0}=c^2 Y_{\ell_2}\,,
\qquad\qquad  
Y_{\ell_1} = s^2 Y_{\ell_2}\,,  
\end{equation}
where $c^2\equiv |\langle \ell_0|\ell_2\rangle|^2$ and $s^2=1-c^2$. The
crucial point here is that in general we expect $c^2\neq0$, and since $\ell_0$
is orthogonal to $\ell_1$, $Y_{\ell_0}$ is protected against $N_1$ washouts.
Then a finite part of the asymmetry $Y_{\ell_2}$ from $N_2$ decays survives
through $N_1$ leptogenesis.  A more detailed study~\cite{Engelhard:2006yg}
reveals also some unexpected features.  For example, $Y_{\ell_1}$ is not
driven to zero in spite of the strong $N_1$-related washouts, rather, only the
sum of $Y_{\ell_1}$ and of the Higgs asymmetry $Y_H$ vanishes, but not the two
separately.  (This can be traced back to the presence of a conserved charge
related to $Y_{\ell_0}$.) As a result we obtain $Y_{\ell_1}=-Y_{\ell_0}/4$ and
$Y_H=Y_{\ell_0}/2$, while the total lepton asymmetry is $
Y_L=(3/2)Y_{\ell_0}=(3/2)\,c^2\,Y_{\ell_2}$.

For $10^9\lsim M_1 \lsim 10^{12}\,$GeV the $\ell_{2,1}$ flavor structures are
only partially resolved during $N_1$ leptogenesis, and a similar result is
obtained.  However, when $M_1 \lsim 10^9\,$GeV and the full flavor basis
$(\ell_e,\ell_\mu,\ell_\tau)$ is resolved, there are no directions in flavor
space where an asymmetry can remain protected, and then the whole $Y_{\ell_2}$
can be erased.  In conclusion, the common assumption that when $N_1$
leptogenesis occurs in the strong washout regime the final BAU is independent
of initial conditions, does not hold in general, and is justified only in the
following cases~\cite{Engelhard:2006yg}: {\it i)}~Vanishing decay asymmetries
and/or efficiency factors for $N_{2,3}$ ($\epsilon_{{2}}\eta_{2}\approx 0$ and
$\epsilon_{{3}}\eta_{3}\approx 0$); {\it ii)}~$N_1$-related washouts are still
significant at $T\lsim10^9\,$GeV; {\it iii)}~Reheating occurs at a temperature
in between $M_2$ and $M_1$.  In all other cases the $N_{2,3}$-related
parameters cannot be ignored when calculating the BAU, and any constraint
inferred from analyses based only on $N_1$ leptogenesis are not reliable.


\begin{theacknowledgments}
  I thank Guy Engelhard, Yuval Grossman, Yosef Nir, Juan Racker and Esteban
  Roulet for the enjoyable collaboration, and Diego Aristizabal, Marta Losada,
  Luis Alfredo Mu\~noz, Jorge Nore\~na and Antonio Riotto for useful
  discussions.  This work was supported in part by INFN in Italy and by
  Colciencias in Colombia.
\end{theacknowledgments}








\end{document}